\documentclass[12pt,leqno]{article}
\usepackage{pdflscape}
\usepackage[justification=centering]{caption}
\usepackage{ amssymb }
\usepackage{amsmath}
\usepackage{amsfonts}
\newtheorem{thm}{Theorem}
\newtheorem{prop}[thm]{Proposition}
\newtheorem{lem}[thm]{Lemma}

\newcommand{\pf}{\noindent {\bf Proof.  }}
\newcommand{\qed}{\hfill $\Box$ \\}

\font\msbm=msbm10 at 12pt
\newcommand{\Z}{\mbox{\msbm Z}}

\begin{document}
\title{ On self-dual double circulant codes}
\author{Adel Alahmadi\thanks{Math Dept, King Abdulaziz University, Jeddah, Saudi Arabia, {Email: \tt adelnife2@yahoo.com}},
Funda \"{O}zdemir\thanks{Sabanc\i \ University, FENS, 34956, \.{I}stanbul, Turkey. {Email: \tt fundaeksi@sabanciuniv.edu}},
Patrick Sol\'e\thanks{ CNRS/LTCI, Telecom ParisTech, Universit\'e de Paris-Saclay, 75 013 Paris, France \& King Abdulaziz University, Math. Dept, Jeddah, KSA.
{Email: \tt sole@enst.fr}}
}
\date{}
\maketitle

\begin{abstract} Self-dual double circulant codes of odd dimension are shown to be dihedral in even characteristic and consta-dihedral in odd characteristic.
Exact counting formulae are derived for them and used to show they
contain families of codes with relative distance satisfying a modified Gilbert-Varshamov bound.
\end{abstract}

\vspace*{1cm}

\bf Key Words\rm : quasi-cyclic codes, dihedral group, consta-dihedral codes, Artin primitive root conjecture

\section{Introduction} It has been known for forty years that quasi-cyclic codes are good \cite{CPW} in the asymptotic sense: the product of their rate by their relative distance does
not vanish when the length goes to infinity. 
Thirteen years ago, it was shown that even the self-dual subclass was good \cite{LS2}. A decade ago,
it was proved that binary dihedral codes of rate one half were good \cite{BM}, then that their self-dual doubly even subclass is also good \cite{W}. The last two papers used
a non-constructive probabilistic argument, where the order of $2$
modulo the length is controlled but not determined. In the present article, we will consider so-called pure double circulant codes, that is 2-quasi cyclic codes with a systematic
generator matrix consisting of two circulant matrices. These codes have been studied in a number of papers since the 1960's \cite{CPW,K,VR}. 
In particular, it is known that binary extended square codes, which form
one of the oldest and most studied family of self-dual codes, are double circulant in many lengths
\cite{J,MM}. We will show that double circulant self-dual codes over an arbitrary finite field of order $q$ are either dihedral or consta-dihedral
depending on the parity of $q.$ (A special case of the first statement is anticipated in \cite{MM}). While the notion of dihedral codes has been considered 
by several authors, the notion of constadihedral codes has been introduced in \cite{VR} in terms of twisted group rings. We give an alternative definition
in terms of group representations. We believe, but do not prove here that the two definitions are related.

Further, building on the Chinese Remainder Theorem (CRT) approach of \cite{LS}, we will give exact counting formulae for these codes. From there, we will
give an alternative proof that dihedral codes are good, with codes of lengths a prime number as in \cite{CPW}. Our proof depends on 
Artin conjecture \cite{M}, proved under the Generalized Riemann Hypothesis 
(GRH) in \cite{H}. It is however, conceptually clearer, and valid for more general alphabets than that of \cite{BM}.
Also we give a new family of good long self-dual
quasi-cyclic codes. They differ from that of \cite{LS2} by the index, the power of the shift under which they are invariant.

The material is organized as follows. Section 2 collects together the definitions and notation that we need thereafter. Section 3 studies the automorphism group of self-dual double circulant
codes, first for even then for odd characteristic. A general notion of consta-dihedral codes is introduced in the language of representation theory. Section 4 studies the asymptotics of double circulant self-dual codes
by combining enumerative formulae with the expurgated random coding argument made familiar by the Gilbert-Varshamov bound.
\section{Definitions and notation}
Let $GF(q)$  denote a finite field of characteristic $p.$
In the following, we will consider codes over $GF(q)$ of length $2n$ with $n$ odd and coprime to $q.$ Their generator matrix $G$ will be of the form
$G=(I,A)$ where $I$ is the identity matrix of order $n$ and $A$ is a circulant matrix of the same order. We will call these codes {\em double circulant}.
These codes are sometimes called pure double circulant to distinguish them 
from bordered double circulant which are not quasi-cyclic \cite{T+}.

By a {\em dihedral} group $D_n,$ we will denote the group of order $2n$ with two generators
$r$ and $s$ of respective orders $n$ and $2$
and satisfying the relation $srs=r^{-1}.$
A code of length $2n$ is called {\em dihedral} if it is invariant under $D_n$ acting transitively on its coordinate places.

If $C(n)$ is a family of codes of parameters $[n,k_n,d_n],$ the rate $R$ and relative distance $\delta$ are defined as
$$R=\limsup_{n \rightarrow \infty}\frac{k_n}{n},$$
and
$$\delta=\liminf_{n \rightarrow \infty}\frac{d_n}{n}.$$
Both limits are finite as limits of bounded quantities.
Such a family of codes is said to be {\it good } if $R\delta \neq 0.$
\section{Symmetry}
\subsection{Even $q$}
Let $q$ be an even prime power. Let $M_n(q)$ denotes the set of all $n\times n$ matrices over $GF(q)$. 

\begin{lem}\label{iso}
If $A$ is a circulant matrix in $M_n(q),$ then there exists an $(n\times n)$-permutation matrix $P$ such that $PAP=A^{t}$ where $A^{t}$ denotes the transpose of $A$.
\end{lem}

\pf Assume, for simplicity, that $n$ is odd. Denote by $\pi$ the permutation $(1,n)(2,n-1)\cdots (\frac{n-1}{2},\frac{n+3}{2}).$ 
Permuting the columns of $A$ with respect to $\pi$ yields a symmetric back-circulant matrix. Let $P$ be the permutation matrix attached to $\pi.$ The preceding explanation shows
that $AP=(AP)^t=P^tA^t,$ or $PAP=A^t.$

\qed

\begin{thm}\label{dihedral}
For $n\geq 3$ odd, and $q$ even, every self-dual double circulant code over $GF(q)$ of length $2n$ is dihedral.
\end{thm}

\pf Let $C$ be a self-dual double circulant code of length $2n$ with generator matrix $G=(I,A)$. The parity-check matrix $H=(A^{t},I)$ is also a generator matrix of $C$ 
due to self-duality. Let $P$ be the $(n\times n)$ permutation matrix such that $PAP=A^{t}$. Since left multiplication by $P$ amounts to changing the positions of some rows, 
$PH=(PA^{t},P)$ is also a generator matrix for $C$.

On the other hand, right multiplication of $PH$ by $P$ is equivalent to multiplying $PH$ by the block diagonal matrix 
$\left( \begin{smallmatrix} P & 0 \\ 0 & P \end{smallmatrix} \right)$, yielding $PHP=(A,I)$. This right multiplication corresponds to applying the following permutation in $S_{2n}$
$$\pi = (2,n)(3, n-1)\ldots (\frac{n+1}{2},\frac{n+3}{2})(n+2,2n)(n+3,2n-1)\ldots (\frac{3n+1}{2})(\frac{3n+3}{2}).$$
i.e. $PHP=PH\pi$. Moreover, we can obtain the generator matrix $(I,A)$ from $(A,I)$ by applying the permutation
$$\sigma= (1,n+1)(2,n+2)(3,n+3)\ldots (n,2n).$$
Hence $C$ is invariant under the following product
$$\pi\sigma=(1, n+1)(2,2n)(3,2n-1)\ldots(n-1,n+3)(n,n+2).$$
Furthermore, since $I$ and $A$ are circulant matrices, $C$ is invariant under also the permutation
$$\tau=(1,2,\ldots, n)(n+1,n+2,\ldots, 2n).$$
Therefore, $C$ is invariant under the subgroup $\langle \tau, \pi\sigma \rangle$ of $S_{2n}$.\
Since $\tau$ is a product of $n$-cycles and $\pi\sigma$ is a product of transpositions, we have $\tau^n=1$ and $(\pi\sigma)^{2}=1$. Observe that 
$$(\pi\sigma)\tau=(1,n+2)(2,n+1)(3,2n)(4,2n-1)\ldots(n-1,n+4)(n,n+3).$$
Then we can easily obtain the following equality
$$(\pi\sigma)\tau(\pi\sigma)=(1,n,n-1,n-2,\ldots,2)(n+1,2n,2n-1,\ldots,n+2)=\tau^{-1}.$$
Therefore, $\langle \tau, \pi\sigma \rangle$ is isomorphic to the dihedral group $D_n$. \qed
\subsection{Odd $q$}
Recall that a {\em monomial} matrix over $GF(q)$ of order $g$ has exactly one nonzero element per row and per column. 
The monomial matrices form a group $M(g,q)$ of order $g!(q-1)^g$ under multiplication.  This group is abstractly isomorphic to the wreath product $\Z_{q-1}\wr S_g.$

By a {\em monomial representation} of a group $G$ over $GF(q)$ we shall mean a group morphism from $G$ into $M(g,q).$
A code of length $2n$ will be said to be {\em consta-dihedral} if it is held invariant under right multiplication by a monomial representation of the dihedral group $D_n.$
An alternative, but related definition can be found in \cite{SR}.
We can now state the main result of this subsection.
\begin{thm}\label{consta}
For $n\geq 3$ odd, and $q$ odd, every self-dual double circulant code $C$ of length $2n$ over $GF(q)$ is consta-dihedral.
\end{thm}

\pf Keep the matrix notations of Theorem 2. Let the generator matrix of $C$ be $G=(I,A)$ with $A$ circulant and $AA^t=-I.$ Computing $A^tG=(A^t,-I)$ and 
conjugating by $P$ of Lemma \ref{iso} we get
$PA^tGP=(A,-I).$ Define the antiswap involution $a$ by the rule $a(x,y)=a(y,-x),$ where $x,y$ are vectors of length $n$ over $GF(q).$ Note that $a^2=-1.$ Clearly $ a \in M(2n,q).$ 
Thus $\pi a \in M(2n,q)$ and it preserves $C.$ A monomial representation of $D_n$ is then $\langle \tau , \pi a\rangle.$ Thus $C$ is consta-dihedral.
\qed

\section{Asymptotics}
\subsection{Enumeration}
In this section we give enumerative results for self-dual double circulant codes. It is important to notice that there are 2-quasi-cyclic codes that are not double circulant.
An example in length $168$ is given in \cite{J}. Thus, the formula of \cite[Prop. 6.2]{LS} does not apply.
 
 We will need the following counting formula. An alternative proof for $q$ prime can be found in \cite[Th 1.3, Th 1.3']{Mac} where the number of orthogonal circulant matrices
 over $GF(q)$ for $q$ prime is computed.
 Recall that $-1$ is a square in $GF(q)$, a field of characteristic $p$, if one of the following conditions holds
\begin{enumerate}
 \item $q$ is even
 \item $p \equiv 1 \pmod{4}$
 \item $p \equiv 3 \pmod{4}$ and $q$ is a square.
\end{enumerate}

Note that \cite[Prop. 6.2]{LS} we know that 2-quasi-cyclic self-dual codes, hence a fortiori self-dual double circulant codes over $GF(q)$ exist only if $-1$ is a square in $GF(q).$

{\lem \label{count}Let $n$ denote a positive odd integer. Assume that $-1$ is a square in $GF(q).$ If $x^n-1$ factors as a product of two irreducible polynomials over $GF(q),$ the number of self-dual double circulant codes of length $2n$ is $2(q^\frac{n-1}{2} +1)$
if $q$ is odd and $(q^\frac{n-1}{2} +1)$ if $q$ is even.
}

\pf
By the CRT approach of \cite{LS} any 2-quasi-cyclic code of length $2n$ over $GF(q)$ decomposes as the 'CRT product' of a self-dual code ${\bf C}_1$ of length $2$ over $GF(q)$ 
and of a hermitian self-dual code ${\bf C}_n$ of length $2$
over $GF(q^{n-1}).$ To obtain a double-circulant code we must ensure that the leftmost entry of their generator matrix $G$ is  $G_{1,1}=1.$

If $q$ is even the only possibility  for ${\bf C}_1$ is the code spanned by $[1,1].$ If $q$ is odd there are two codes $[1,a]$ and $[1,-a]$ where $a^2=-1.$

For ${\bf C}_n$ the generator matrix is $[1,b]$ with $b$ such that $1+b^{1+r}=0,$ with $q^{n-1}=r^2.$ 
By finite field theory, this equation in $b$ admits $1+r$ roots in $GF(r^2).$ Note that if $q$ is even, $b$ ranges over the elements of order dividing $1+r=\frac{r^2-1}{r-1},$
and that if $q$ is odd, $b^2$ ranges over elements of order $2(1+r).$ In both cases, we use the fact that the multiplicative group of $GF(r^2)$ is cyclic of order $r^2-1.$
\qed

The following, more general, result is an analogue for double circulant codes  of the Proposition \cite[Prop.6.2]{LS} for 2-quasi-cyclic codes. 
It is of interest in its own right, but not needed for the asymptotic bounds of this section.

{\prop Let $n$ be an odd integer, and $q$ a prime power coprime with $n.$ Suppose that $-1$ is a square in $GF(q).$ Assume that the factorization of $x^n-1$ into irreducible polynomials over $GF(q)$  is of the form
$$x^n-1=\alpha (x-1)\prod_{j=2}^{s}g_j(x) \prod_{j=1}^{t}h_j(x)h_j^*(x),$$
with $\alpha$ a scalar of $GF(q),$ $n=s+2t$ and $g_j$ a self-reciprocal polynomial of degree $2d_j,$ the polynomial $h_j$ is of degree $e_j$ and $*$ denotes reciprocation.
For convenience, let $g_1=x-1$ and, in case of $n$ even, let $g_2=x+1.$
The number of self-dual 2-quasi-cyclic codes over $GF(q)$ is then
$$4 \prod_{j=3}^s(1+q^{d_j}) \prod_{j=1}^t(q^{e_j}-1)$$ if $q$ is odd and $n$ is even
$$2\prod_{j=2}^s(1+q^{d_j}) \prod_{j=1}^t(q^{e_j}-1) $$ if  $q$ is odd and $n$ is odd
$$\prod_{j=2}^s(1+q^{d_j})  \prod_{j=1}^t(q^{e_j}-1) $$ if  $q$ is even and $n$ is odd.
}

\pf (sketch). The part of the proof dealing with  self-reciprocal polynomials $g_j$ is analogous to the previous lemma. In the case of reciprocal pairs $(h_j,h_j^*)$, note that the number of linear codes of length $2$ over some
$GF(Q)$ admitting, along with their duals, a systematic form is $Q-1,$ all of dimension $1.$ Indeed their generator matrix is of the form $[1,u]$ with $u$ nonzero.
We conclude by letting $Q=q^{e_j}.$
\qed
\subsection{Arithmetic}
In number theory, Artin's conjecture on primitive roots states that a given integer $q$ which is neither a perfect square nor $-1$ 
is a primitive root modulo infinitely many primes $\ell$ \cite{M}. It was proved conditionally under GRH by Hooley \cite{H}. 
In this case, by the correspondence between cyclotomic cosets and irreducible factors of  $x^\ell-1$ \cite{HP}, the factorization of $x^\ell-1$
into irreducible polynomials over $GF(q)$ contains exactly two factors, one of which is $x-1$ \cite{CPW}.
\subsection{Distance bound}
We will need a $q$-ary version of a classical lemma from \cite{CPW}. Let $a(x)$ denote a polynomial of $GF(q)[x]$ coprime with $x^n-1,$ and let $C_a$ be the double circulant code with generator matrix
$(1,a).$ Assume the factorization of $x^n-1$ into irreducible polynomials is $x^n-1=(x-1)h(x).$ We call {\it constant vectors} the codewords of the cyclic code of length $n$
generated by $h.$

{\lem \label{CPW}If $u$ is not a constant vector then there are only at most $(q-1)$ polynomials $a$ such that $u\in C_a.$}

\pf
Write $u=(v,w)$ with $v,w$ of length $n.$ The condition $u \in C_a$ is equivalent to the equation $w=av \pmod{x^n-1}.$ If $v$ is invertible$\pmod{x^n-1},$ then $v$ is uniquely
determined by this equation. If not and if $u$ is not a constant vector the only possibility is that both $w$ and $v$ are multiples of $(x-1).$
Letting $v=(x-1)v',$ and $w=(x-1)w',$ yields $w'=av'\pmod{h(x)},$ which gives $a$ $\pmod{h(x)},$ since $v'$ is invertible $\pmod{h(x)}.$ 
Now $a \pmod{(x-1)}$ can take $q-1$ nonzero values.
The result follows by the CRT applied to $a,$ since $a,$ being of degree at most $n-1$ is completely determined by its residue $\pmod{x^n-1}.$
\qed

Recall the $q-$ary entropy function defined for $0<x< \frac{q-1}{q}$ by $$ H_q(x)=x\log_q(q-1-x\log_q(x)-(1-x)\log_q(1-x).$$
We are now ready for the main result of this section.

{\thm If $q$ is not a square, then there are infinite families of self-dual double circulant codes of relative distance $$\delta \ge H_q^{-1}(\frac{1}{4}).$$}

\pf
Let $q$ be fixed and $n$ a prime going to infinity that satisfies the Artin conjecture for $q$ . The double circulant codes containing a vector of weight $d\sim \delta n$ or less are by 
standard entropic estimates of \cite{HP}
and Lemma \ref{CPW} of the order of $(q-1) q^{2n H_q(\delta)},$ up to subexponential terms. This number will be less than the total number of self-dual double circulant codes
which is by Lemma \ref{count} of the order of $q^{n/2},$ as soon as $\delta$ is of the order of the stated bound.
\qed
\section{Conclusion and Open problems}
In this paper, we have studied the class of double circulant self-dual codes over finite fields, under the aspects of symmetry, enumeration, and asymptotic performance. The self-dual condition
shows that these codes in odd dimension are held invariant by the dihedral group of order the length of the code in the even characteristic case,
and by a monomial representation of that group
in the odd characteristic case. It is possible that a similar phenomenon occurs for $n$ even and, more generally, for quasi-cyclic codes of higher index than two. Further, we have derived an exact enumeration
formula for this family of codes. This formula can be interpreted as an enumeration of circulant orthogonal matrices over finite fields,
thus generalizing a result of MacWilliams \cite{Mac}
in the prime field case, to general finite fields. Our approach to asymptotic bounds on the minimum distance relies on some deep number-theoretic conjectures ( Artin or GRH). 
It would be a worthwhile
task to remove this dependency by looking at lengths where the factorization of $x^n-1$ into irreducible polynomials contains more than two elements.

{\bf Acknowledgement:} The authors are indebted to Hatoon Shoaib for helpful discussions. The second author was supported by T\"{U}B\.{I}TAK 2214-International Doctoral Research Fellowship Programme.

\end{document}